\documentclass[english]{article}
\usepackage{graphicx}
\usepackage[T1]{fontenc}
\usepackage[latin1]{inputenc}
\usepackage{geometry}
\usepackage{amsmath}
\usepackage{amssymb}

\makeatletter

\providecommand{\LyX}{L\kern-.1667em\lower.25em\hbox{Y}\kern-.125emX\@}

\newcounter{draft}
\usepackage{amssymb} 
\usepackage{ifthen}
\ifthenelse{\thedraft=1}{\usepackage[notref,notcite]{showkeys}}{}

\usepackage{hyperref}
\usepackage{cite} 

\numberwithin{equation}{section}

\newcommand{\eref}{\def\theequation{{\bf \thesection}.\arabic{equation}}}

\newcommand{\rem}[1]{\ifthenelse{\thedraft=1}{\textsc{\texttt{<< #1 >>}}}{}}

\newcommand{\bmul}{\begin{multicols}{2}[\vspace{-1cm}]}
\newcommand{\emul}{\end{multicols}\vspace{-0.5cm}}

\date{\rem{\today}}
\newcommand{\Teil}[2]{#2} 

\usepackage{babel}
\makeatother
\begin{document}
\Teil{Macros}{
\newcommand{\bs}[1]{\boldsymbol{#1}}

\newcommand{\mf}[1]{\mathfrak{#1} }

\newcommand{\mc}[1]{\mathcal{#1}}

\newcommand{\norm}[1]{{\parallel#1 \parallel}}

\newcommand{\Norm}[1]{\left\Vert #1 \right\Vert }

\newcommand{\partiell}[2]{\frac{\partial#1 }{\partial#2 }}

\newcommand{\Partiell}[2]{\left( \frac{\partial#1 }{\partial#2 }\right) }

\newcommand{\ola}[1]{\overleftarrow{#1}}

\newcommand{\lpartial}{\overleftarrow{\partial}}

\newcommand{\partl}[1]{\frac{\partial}{\partial#1}}

\newcommand{\partr}[1]{\frac{\lpartial}{\partial#1}}

\newcommand{\funktional}[2]{\frac{\delta#1 }{\delta#2 }}

\newcommand{\funktl}[1]{\frac{\delta}{\delta#1}}

\newcommand{\funktr}[1]{\frac{\ola{\delta}}{\delta#1}}

\newcommand{\de}{{\bf d}\!}

\newcommand{\es}{{\bf s}\!}

\newcommand{\dew}{\bs{d}^{\textrm{w}}\!}

\newcommand{\Lie}{\bs{\mc{L}}}

\newcommand{\Dorf}{\bs{\mc{D}}}

\newcommand{\pe}{\bs{\partial}}

\newcommand{\De}{\textrm{D}\!}

\newcommand{\total}[2]{\frac{\de#1 }{\de#2 }}

\newcommand{\Frac}[2]{\left( \frac{#1 }{#2 }\right) }

\newcommand{\To}{\rightarrow}
 
\newcommand{\ket}[1]{|#1 >}

\newcommand{\bra}[1]{<#1 |}

\newcommand{\Ket}[1]{\left| #1 \right\rangle }

\newcommand{\Bra}[1]{\left\langle #1 \right| }
 
\newcommand{\braket}[2]{<#1 |#2 >}

\newcommand{\Braket}[2]{\Bra{#1 }\left. #2 \right\rangle }

\newcommand{\kom}[2]{[#1 ,#2 ]}

\newcommand{\Kom}[2]{\left[ #1 ,#2 \right] }

\newcommand{\abs}[1]{\mid#1 \mid}

\newcommand{\Abs}[1]{\left| #1 \right| }

\newcommand{\erw}[1]{\langle#1\rangle}

\newcommand{\Erw}[1]{\left\langle #1 \right\rangle }

\newcommand{\bei}[2]{\left. #1 \right| _{#2 }}

\newcommand{\dann}{\Rightarrow}

\newcommand{\q}[1]{\underline{#1 }}

\newcommand{\hoch}[1]{{}^{#1 }}

\newcommand{\tief}[1]{{}_{#1 }}

\newcommand{\lqn}[1]{\lefteqn{#1}}

\newcommand{\os}[2]{\overset{\lqn{#1}}{#2}}

\newcommand{\us}[2]{\underset{\lqn{{\scriptstyle #2}}}{#1}}

\newcommand{\ous}[3]{\underset{#3}{\os{#1}{#2}}}

\newcommand{\zwek}[2]{\begin{array}{c}
#1\\
#2\end{array}}

\newcommand{\drek}[3]{\begin{array}{c}
#1\\
#2\\
#3\end{array}}

\newcommand{\UB}[2]{\underbrace{#1 }_{\le{#2 }}}

\newcommand{\OB}[2]{\overbrace{#1 }^{\le{#2 }}}

\newcommand{\tr}{\textrm{tr}\,}

\newcommand{\Tr}{\textrm{Tr}\,}

\newcommand{\Det}{\textrm{Det}\,}

\newcommand{\diag}{\textrm{diag}\,}

\newcommand{\Diag}{\textrm{Diag}\,}

\newcommand{\one}{1\!\!1}

\newcommand{\fussend}{\diamond}

\newcommand{\eps}{\varepsilon}

\newcommand{\dali}{\Box}
\newcommand{\choice}[2]{\ifthenelse{\thechoice=1}{#1}{\ifthenelse{\thechoice=2}{#2}{\left\{  \begin{array}{c}

 #1\\
#2\end{array}\right\}  }}}
 
\newcommand{\lchoice}[2]{\ifthenelse{\thechoice=1}{#1}{\ifthenelse{\thechoice=2}{#2}{\left\{  \begin{array}{c}

 #1\\
#2\end{array}\right.}}}
 
\newcommand{\lcsign}{\ifthenelse{\thechoice=1}{+}{\ifthenelse{\thechoice=2}{-}{\pm}}}
 
\newcommand{\lcmsign}{\ifthenelse{\thechoice=1}{-}{\ifthenelse{\thechoice=2}{+}{\mp}}}

\newcommand{\lcconst}{c}
{}
 
\newcommand{\weyl}{\alpha}

\newcommand{\greq}{=_{g}}

\newcommand{\grequiv}{\equiv_{g}}

\newcommand{\grdef}{:=_{g}}

\newcommand{\Greq}{=_{G}}

\newcommand{\Grequiv}{\equiv_{G}}

\newcommand{\Grdef}{:=_{G}}

\newcommand{\Ggreq}{=_{Gg}}

\newcommand{\Greqornot}{=_{(G)}}

\newcommand{\Grequivornot}{\equiv_{(G)}}

\newcommand{\Grdefornot}{:=_{(G)}}

\newcommand{\greqornot}{=_{(g)}}

\newcommand{\grequivornot}{\equiv_{(g)}}

\newcommand{\grdefornot}{:=_{(g)}}

\newcommand{\fatkomma}{\textrm{{\bf ,}}}

\newcommand{\basis}{\boldsymbol{\mf{t}}}

\newcommand{\ip}{\imath}

\newcommand{\Beta}{\textrm{\Large$\beta$}}

\newcommand{\sBeta}{\textrm{\large$\beta$}}

\newcommand{\be}{\bs{b}}

\newcommand{\ce}{\bs{c}}

\newcommand{\Q}{\bs{Q}}

\newcommand{\mm}{\bs{m}\ldots\bs{m}}

\newcommand{\nn}{\bs{n}\ldots\bs{n}}

\newcommand{\kk}{k\ldots k}

\newcommand{\OO}{\bs{\Omega}}

\newcommand{\oo}{\bs{o}}

\newcommand{\tet}{\bs{\theta}}

\newcommand{\Es}{\bs{S}}

\newcommand{\Ce}{\bs{C}}

\newcommand{\lam}{\bs{\lambda}}

\newcommand{\ro}{\bs{\rho}}

\newcommand{\cov}{\textrm{D}_{\tet}}

\newcommand{\feps}{\bs{\eps}}

\newcommand{\qu}{\textrm{Q}_{\tet}}

\newcommand{\dimw}{d_{\textrm{w}}}

\newcommand{\mteta}{\mu(\tet)}

\newcommand{\msig}{d^{\lqn{\hoch{\dimw}}}\sigma}

\newcommand{\msigp}{d^{\hoch{{\scriptscriptstyle \dimw}\lqn{{\scriptscriptstyle -1}}}}\sigma}

\newcommand{\backtilde}{\!\!\tilde{}\,\,}

\title{A radia\c{c}\~ao de corpo negro e o surgimento da f\'isica qu\^antica}

\author{\begin{picture}(0,0)\unitlength=1mm\put(80,40){Notas de aula}\put(80,35){F\'isica contempor\^anea I}
\end{picture} \textbf{Tiago Carvalho Martins}%
\thanks{tiagocm@ufpa.br%
}\emph{\vspace{.5cm}} \emph{}\\
\emph{Faculdade de F\'isica, Universidade Federal do Sul e Sudeste do Par\'a, UNIFESSPA}\\
\emph{68505-080, PA, Brazil\vspace{.5cm}} \emph{}\\}

\maketitle
\begin{abstract}
It is shown how the study of blackbody radiation in the early twentieth century by the German physicist Max Planck 
gave rise to the quantum theory. \noindent \vspace{2cm} \\
\centering{\textbf{Resumo}} \\
\'E mostrado como o estudo da radia\c{c}\~ao de corpo negro no in\'icio do s\'eculo XX pelo f\'isico alem\~ao Max Planck 
deu origem \`a teoria qu\^antica. \noindent \vspace{2cm}
\end{abstract}
}

\Teil{A}{
\newpage 

\tableofcontents{}\newpage\eref

\section{Introdu\c{c}\~ao}

Em 1901, Max Planck publicou um trabalho sobre a lei de radia\c{c}\~ao de corpos negros, em que prop\~oe a hip\'otese de que a energia
\'e quantizada \cite{Artigo.Planck:1901}. Esse trabalho \'e considerado o marco inicial no surgimento da f\'isica qu\^antica
\cite {Livro.Eisberg:1985}.

A \textit{radia\c{c}\~ao t\'ermica} \'e aquela emitida ou absorvida por um corpo devido \`a sua temperatura.
Num corpo mais quente que o meio, a \textit{taxa de radia\c{c}\~ao} supera a 
\textit{taxa de absor\c{c}\~ao}.
As taxas de absor\c{c}\~ao e de radia\c{c}\~ao tornam-se iguais quando o corpo atinge o \textit{equil\'ibrio t\'ermico}.

Em geral, a quantidade de energia liberada, em uma certa temperatura, para uma dada faixa de frequ\^encias, 
depende das propriedade f\'isicas do radiador. O \textit{corpo negro} \'e uma exce\c{c}\~ao dessa regra. 
 
O \textit{corpo negro} \'e o absorvedor t\'ermico por excel\^encia, e portanto, de acordo com a
\textit{lei de Kirchhoff da radia\c{c}\~ao}, o melhor radiador.

Ao longo do texto, a express\~ao \textit{"corpo negro"} ser\'a utilizada para denominar 
um pequeno orif\'icio em uma cavidade fechada, pois ele se constitui em um excelente absorvedor, haja vista  
a radia\c{c}\~ao que o atravessa ser refletida v\'arias vezes no interior da cavidade, 
sendo continuamente absorvida pelas paredes internas. Isso \'e ilustrado na figura \ref{FiguraOrificioCavidade}.

\begin{figure}[!htb]
\centering
\includegraphics[scale=0.3]{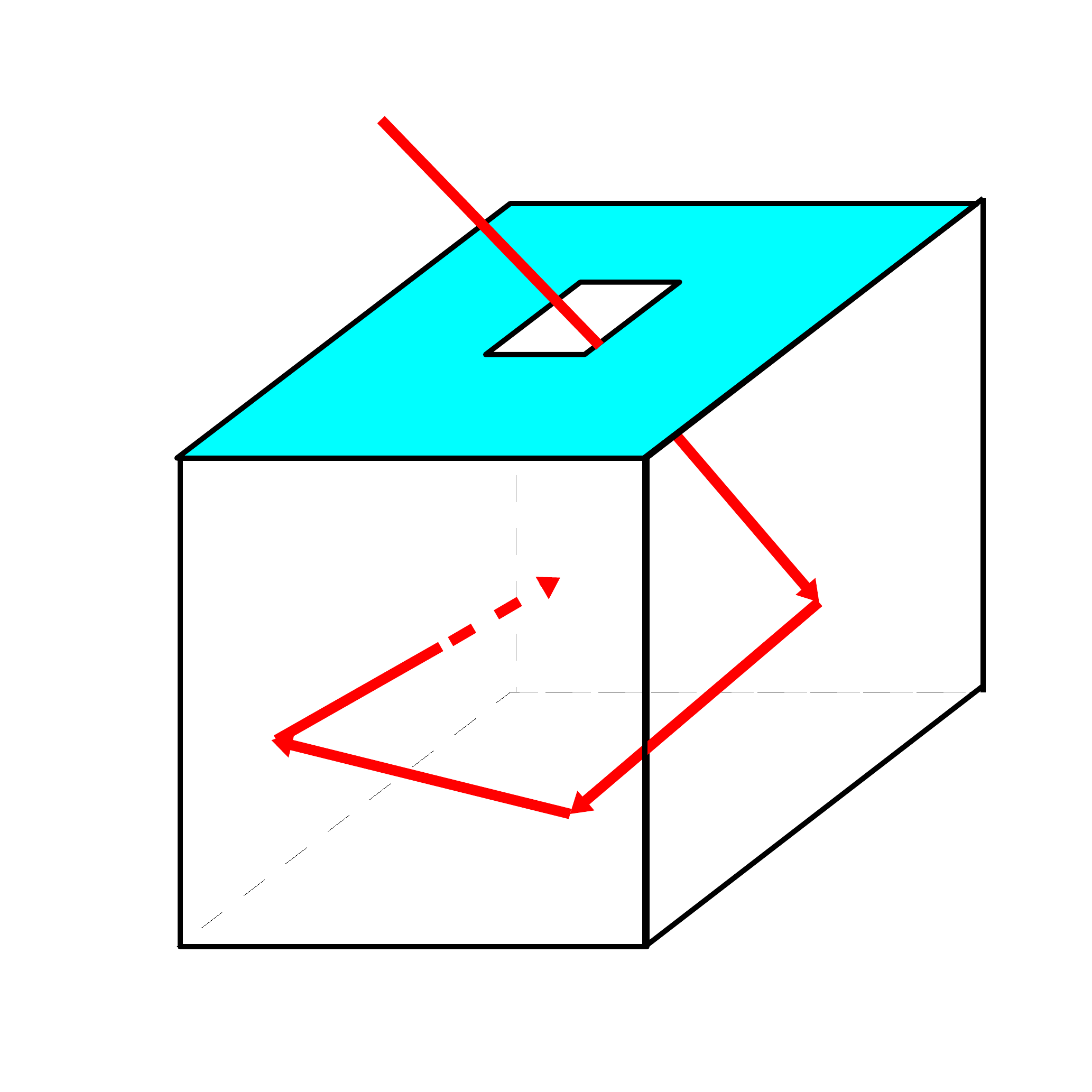}
\caption{A radia\c{c}\~ao que atravessa o orif\'icio \'e refletida v\'arias vezes no interior da cavidade, 
sendo continuamente absorvida pelas paredes internas.}
\label{FiguraOrificioCavidade}
\end{figure}

\section{Alguns conceitos importantes}

\label{SectionConceitosPreliminares} 

A \textit{radi\^ancia por unidade de intervalo de frequ\^encia} $R_T(f)$ est\'a relacionada a quantidade $R_T(f)df$, 
que \'e a energia t\'ermica irradiada pelo objeto por unidade de \'area e tempo no intervalo de frequ\^encias de $f$ a $f+df$.

A \textit{radi\^ancia espectral} $R_T$ \'e a energia t\'ermica total (em todo o espectro de frequ\^encias)
irradiada pelo objeto por unidade de \'area e tempo:

\begin{equation}
R_T=\int_0^\infty R_T(f)df
\label{RT}
\end{equation}

Para um corpo negro vale a \textit{lei do deslocamento de Wien}:

\begin{equation}
\lambda_{max} T = 2,898 \times 10^{-3} m.K
\label{leiwien}
\end{equation} onde $\lambda_{max}$ \'e o comprimento de onda para o qual a radi\^ancia \'e m\'axima e $2,898 \times 10^{-3} m.K$
\'e a \textit{constante de Wien}.

Para um corpo negro vale tamb\'em a \textit{lei de Stefan-Boltzmann}:

\begin{equation}
R_T = \sigma T^4
\label{StefanBoltzmann}
\end{equation} onde $\sigma=5,6704 \times 10^{-8} W m^{-2} K^{-4}$ \'e a \textit{constante de Stefan-Boltzmann}.

O sol tem $\lambda_{max}=510~nm$, utilizando (\ref{leiwien}), podemos estimar que a temperatura do sol \'e de $5682 K$.
Al\'em disso, utilizando (\ref{StefanBoltzmann}), podemos estimar que a densidade superficial de pot\^encia irradiada pelo sol
\'e de $5910,4 W/cm^2$. 

Existe uma proximidade de $\lambda_{max}=510~nm$ do sol com os picos de sensibilidade visual humana diurno e noturno, 
respectivamente, $560~nm$ e $510~nm$, devida ao processo de \textit{sele\c{c}\~ao natural} ao longo da evolu\c{c}\~ao da esp\'ecie
humana.

A \textit{densidade de energia} $\rho_T(f)df$ \'e a energia contida em um volume unit\'ario da cavidade a temperatura $T$ 
no intervalo de frequ\^encias de $f$ a $f+df$.

Um orif\'icio em uma cavidade irradia mais intensamente na faixa de frequ\^encias em que a densidade de energia dentro da cavidade
\'e maior (ver ap\^endice \ref{sectionrelacaodensidadeenergia}):

\begin{equation}
R_T(f)df \propto \rho_T(f)df 
\label{RTprop}
\end{equation}

Dada a energia m\'edia $<E>$, a densidade de energia \'e dada por (ver ap\^endice \ref{sectionnumerodemodos}):
 
\begin{equation}
\rho_T(f)df=\frac{8\pi f^2}{c^3}<E> df
\label{densidadedeenergia}
\end{equation}

\section{Energia m\'edia para o caso cont\'inuo}

A energia m\'edia $<E>$ para o caso discreto \'e (ver ap\^endice \ref{calculoenergiamedia}):

\begin{equation}
<E> = \frac{\sum_{i=0}^\infty E_iP_i}{\sum_{i=0}^\infty P_i}=
\frac{\sum_{i=0}^\infty E_i(B e^{-\frac{E_i}{kT}})}{\sum_{i=0}^\infty B e^{-\frac{E_i}{kT}}}=
\frac{\sum_{i=0}^\infty E_i e^{-\frac{E_i}{kT}}}{\sum_{i=0}^\infty e^{-\frac{E_i}{kT}}}
\label{energiamedia}
\end{equation}

Passando para o caso cont\'inuo fica:

\begin{equation}
<E> = \frac{\int_{0}^\infty E e^{-\frac{E}{kT}} dE}{\int_{0}^\infty e^{-\frac{E}{kT}}dE}
\label{energiamedia1a}
\end{equation}

Com a integral do denominador dada por:

\begin{equation}
\int_{0}^\infty e^{-\frac{E}{kT}}dE=-kT\int_{0}^\infty e^{-\frac{E}{kT}}d(-\frac{E}{kT})
=-kT~e^{-\frac{E}{kT}}|_0^\infty=-kT~[\lim_{E\rightarrow \infty}~e^{-\frac{E}{kT}}-e^{0}]
\label{energiamedia2}
\end{equation}

Mas:

\begin{equation}
\lim_{E\rightarrow \infty}~e^{-\frac{E}{kT}}=0
\label{energiamedia3}
\end{equation}

Portanto:

\begin{equation}
\int_{0}^\infty e^{-\frac{E}{kT}}dE=-kT~[0-1]=kT
\label{energiamedia4}
\end{equation}

A integral do numerador pode ser resolvida por integra\c{c}\~ao por partes fazendo $p=E$ e $dq=e^{-\frac{E}{kT}}dE$, e tendo
$dp=1$ e $q=-kTe^{-\frac{E}{kT}}$:

\begin{equation}
\int_{0}^\infty pdq = pq|_0^\infty-\int_{0}^\infty dp q
\end{equation}

\begin{equation}
\int_{0}^\infty Ee^{-\frac{E}{kT}} = E(-kTe^{-\frac{E}{kT}})|_0^\infty-\int_{0}^\infty 1 \times (-kTe^{-\frac{E}{kT}}dE)
= -kT(Ee^{-\frac{E}{kT}}|_0^\infty-\int_{0}^\infty e^{-\frac{E}{kT}}dE)
\label{energiamedia5a}
\end{equation}

Mas:

\begin{equation}
\lim_{E\rightarrow \infty}~Ee^{-\frac{E}{kT}}=\lim_{E\rightarrow \infty}~\frac{E}{e^{\frac{E}{kT}}}=\frac{\infty}{\infty}
\label{energiamedia6}
\end{equation}

Pela segunda regra de L'hospital:

\begin{equation}
\lim_{E\rightarrow \infty}~Ee^{-\frac{E}{kT}}=\lim_{E\rightarrow \infty}~\frac{(E)'}{(e^{\frac{E}{kT}})'}=\lim_{E\rightarrow \infty}~\frac{1}{e^{\frac{E}{kT}}/(kT)}=0
\label{energiamedia7}
\end{equation}

Substituindo (\ref{energiamedia4}) e (\ref{energiamedia7}) em (\ref{energiamedia5a}):

\begin{equation}
\int_{0}^\infty Ee^{-\frac{E}{kT}} = -kT(0-kT)=(kT)^2
\label{energiamedia8}
\end{equation}

Substituindo (\ref{energiamedia4}) e (\ref{energiamedia8}) em (\ref{energiamedia1a}):

\begin{equation}
<E> = \frac{(kT)^2}{kT}=kT
\label{energiamedia1}
\end{equation}

\section{F\'ormula para radia\c{c}\~ao de corpo negro de Rayleigh-Jeans}

Para um n\'umero de estados de enegia cont\'inuo, 
a densidade de energia pode ser obtida pelo substitui\c{c}\~ao de (\ref{energiamedia1}) em (\ref{densidadedeenergia}):

\begin{equation}
\rho_T(f) df= \frac{8\pi f^2}{c^3} kT df
\label{densidadevolumetricacontinua}
\end{equation}

O aumento de $\rho_T(f)df$ produzido na densidade de energia devido ao incremento $df$ corresponde
\`a redu\c{c}\~ao de $\rho_T(\lambda)d\lambda$ na densidade de energia devida ao decremento $d\lambda$:

\begin{equation}
\rho_T(\lambda) d\lambda = - \rho_T(f) df
\label{dfdlambda}
\end{equation}

Como a frequ\^encia \'e $f=\frac{c}{\lambda}$, $df=-\frac{c}{\lambda^2}d\lambda$, portanto:

\begin{equation}
\rho_T(\lambda) = - \rho_T(f) \frac{df}{d\lambda} =  - \rho_T(f) (-\frac{c}{\lambda^2})= \rho_T(\frac{c}{\lambda}) \frac{c}{\lambda^2}
\label{densidadevolumetricacontinua1}
\end{equation}

Logo:

\begin{equation}
\rho_T(\lambda) = \frac{8\pi (\frac{c}{\lambda})^2}{c^3} kT \frac{c}{\lambda^2}=\frac{8\pi}{\lambda^4} kT
\label{densidadevolumetricacontinua2}
\end{equation}

\section{Energia m\'edia para o caso discreto}

Para o caso em que h\'a um n\'umero discreto de estados de energia, tem-se $E_n=n\epsilon$, onde 
$\epsilon$ \'e o quantum de energia, e portanto, o somat\'orio do denominador \'e dado por:

\begin{equation}
\sum_{n=0}^\infty e^{-\frac{E_n}{kT}}=\sum_{n=0}^\infty exp(-\frac{n\epsilon}{kT})
\label{denominadordiscreto}
\end{equation}

O limite da s\'erie geom\'etrica \'e dado por (ver se\c{c}\~ao \ref{series}):

\begin{equation}
\sum_{n=0}^\infty x^n=\frac{1}{1-x}
\label{somaPG}
\end{equation}

E fazendo $x=exp(-\frac{\epsilon}{kT})$:

\begin{equation}
\sum_{n=0}^\infty e^{-\frac{E_n}{kT}}=\sum_{n=0}^\infty x^n=\frac{1}{1-x}=\frac{1}{1-exp(-\frac{\epsilon}{kT})}
\label{denominadordiscreto1}
\end{equation}

O limite da s\'erie $\sum_{n=0}^\infty nx^n$ \'e (ver se\c{c}\~ao \ref{series1}):

\begin{equation}
\sum_{n=0}^\infty nx^n=\frac{x}{(1-x)^2}
\label{somaPG2}
\end{equation}

O somat\'orio do denominador \'e dado por:

\begin{equation}
\sum_{n=0}^\infty E_n e^{-\frac{E_n}{kT}}=\sum_{n=0}^\infty n\epsilon~exp(-\frac{n\epsilon}{kT})
\label{somaPG3}
\end{equation}

Fazendo $x=exp(-\frac{\epsilon}{kT})$:

\begin{equation}
\sum_{n=0}^\infty E_n e^{-\frac{E_n}{kT}}=\epsilon \sum_{n=0}^\infty nx^n = \epsilon \frac{x}{(1-x)^2}=
\epsilon \frac{exp(-\frac{\epsilon}{kT})}{(1-exp(-\frac{\epsilon}{kT}))^2}
\label{somaPG4}
\end{equation}

Substituindo (\ref{somaPG4}) e (\ref{denominadordiscreto1}) em (\ref{energiamedia}):

\begin{equation}
<E>=
\frac{\epsilon~exp(-\frac{\epsilon}{kT})}{1-exp(-\frac{\epsilon}{kT})}
\label{energiamediadiscreta}
\end{equation}

Planck assumiu que
o quantum de energia $\epsilon$ depende da frequ\^encia $f$, $\epsilon=hf$, onde $h=6,626068 \times 10^{-34} m^2 kg / s$ 
\'e a constante de Planck, portanto:

\begin{equation}
<E>=
\frac{hf~exp(-\frac{hf}{kT})}{1-exp(-\frac{hf}{kT})}=
\frac{hf}{(1-exp(-\frac{hf}{kT}))exp(\frac{hf}{kT})}=
\frac{hf}{exp(\frac{hf}{kT})-1}
\label{energiamediadiscreta1}
\end{equation}

\section{F\'ormula para radia\c{c}\~ao de corpo negro de Planck}

A densidade de energia pode ser obtida substituindo (\ref{energiamediadiscreta1}) em (\ref{densidadedeenergia}):

\begin{equation}
\rho_T(f) df=\frac{8\pi h f^3}{c^3} \frac{1}{exp(\frac{hf}{kT})-1}df
\label{densidadevolumetricadiscreta}
\end{equation}

Substituindo (\ref{densidadevolumetricadiscreta}) em (\ref{densidadevolumetricacontinua1}):

\begin{equation}
\rho_T(\lambda)
=\frac{8\pi h (c/\lambda)^3}{c^3} \frac{1}{exp(\frac{h(c/\lambda)}{kT})-1}\frac{c}{\lambda^2}
=\frac{8\pi h c}{\lambda^5} \frac{1}{exp(\frac{hc}{kT\lambda})-1}
\label{densidadevolumetricadiscreta1}
\end{equation}

\section{Compara\c{c}\~ao entre radia\c{c}\~ao de corpo negro de Rayleigh-Jeans e de Planck}

Na figura \ref{fig.planckrayleigh} \'e mostrada a compara\c{c}\~ao entre as f\'ormulas de radia\c{c}\~ao de corpo 
negro de Rayleigh-Jeans e de Planck,
dadas por (\ref{densidadevolumetricacontinua2}) e (\ref{densidadevolumetricadiscreta1}), respectivamente. 
Para pequenos comprimentos de onda, a lei de
Rayleigh-Jeans assume
valores absurdamente grandes (isto \'e conhecido como trag\'edia do ultravioleta). A lei de Planck por sua vez coincide
perfeitamente com os resultados experimentais.

\begin{figure}[!h]
\begin{center}
\includegraphics[scale=1.0, bb = 50 50 400 300]{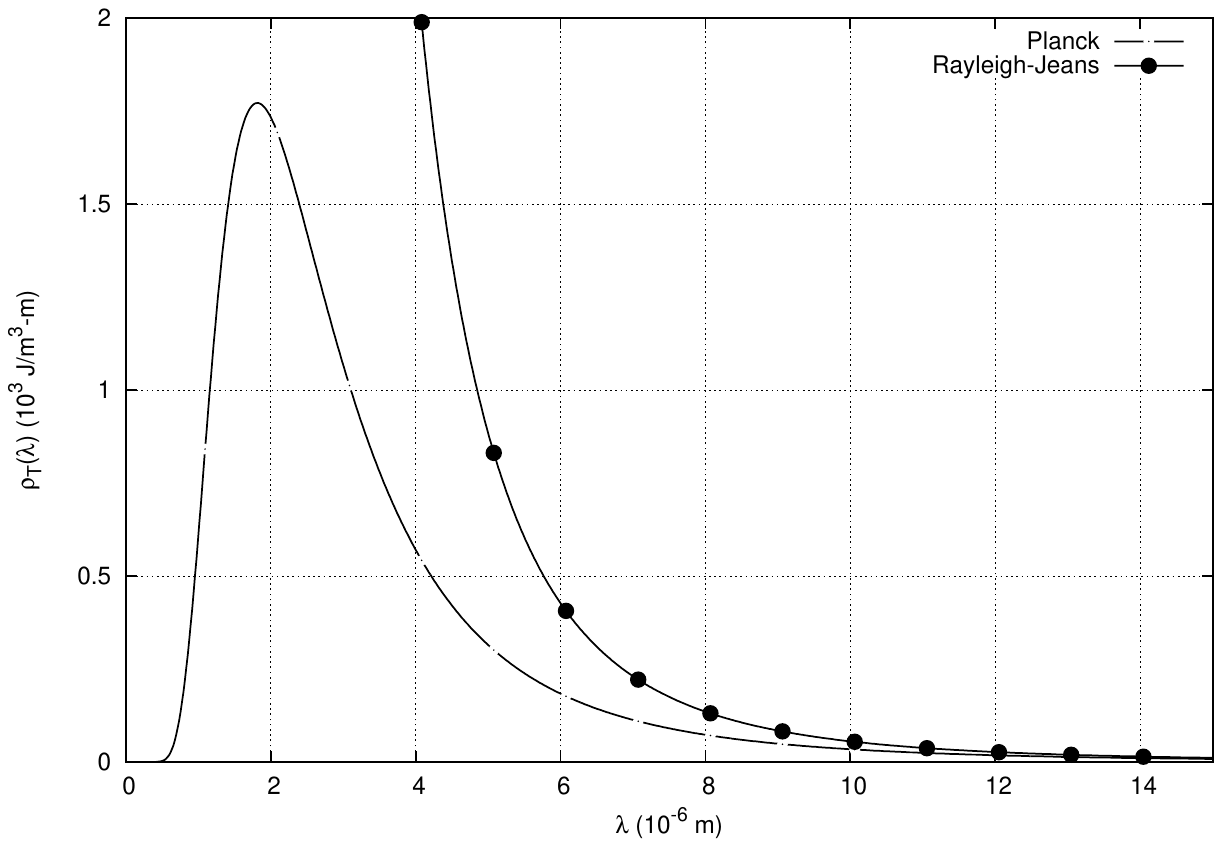}
\end{center}
\caption{Radia\c{c}\~ao de corpo negro de Rayleigh-Jeans e de Planck para a temperatura de 1595 K.}
\label{fig.planckrayleigh}
\end{figure}

\section{Obtendo a lei de Wien a partir da f\'ormula da radia\c{c}\~ao de corpo negro de Planck}

A figura \ref{fig.leidewien}, plotada a partir de (\ref{densidadevolumetricadiscreta1}), ilustra a lei de
Wien mostrada em (\ref{leiwien}), a qual pode ser obtida a partir da
f\'ormula da radia\c{c}\~ao de corpo negro de Planck (\ref{densidadevolumetricadiscreta1}).
Para que a densidade de energia $\rho_T(\lambda)$ seja m\'axima:

\begin{equation}
\frac{d\rho_T(\lambda)}{d\lambda}=0
\label{densidadevolumetricadiscretamaxima}
\end{equation}

Substituindo (\ref{densidadevolumetricadiscreta1}) em (\ref{densidadevolumetricadiscretamaxima}):

\begin{eqnarray}
\frac{d\rho_T(\lambda)}{d\lambda}
=\frac{d}{d\lambda}\left (\frac{8\pi h c}{\lambda^5} \frac{1}{exp(\frac{hc}{kT\lambda})-1} \right )=0 \nonumber \therefore
8\pi h c \frac{d}{d\lambda}\left ( \frac{\lambda^{-5}}{exp(\frac{hc}{kT\lambda})-1} \right )=0 \nonumber \\
\frac{-5\lambda^{-6}(exp(\frac{hc}{kT\lambda})-1)-\lambda^{-5}(exp(\frac{hc}{kT\lambda})-1)'}{(exp(\frac{hc}{kT\lambda})-1)^2} =0 \nonumber \therefore
-5\lambda^{-6}(exp(\frac{hc}{kT\lambda})-1)-\lambda^{-5}exp(\frac{hc}{kT\lambda})\frac{hc}{kT}\frac{-1}{\lambda^2} =0 \nonumber \\
-5\lambda(1-exp(-\frac{hc}{kT\lambda}))+\frac{hc}{kT} =0 \nonumber \therefore
-5(1-exp(-\frac{hc}{kT\lambda}))+\frac{hc}{kT\lambda} =0 
\label{densidadevolumetricadiscreta2}
\end{eqnarray}

Fazendo $x=\frac{hc}{kT\lambda}$:

\begin{equation}
-5(1-e^{-x})+x =0 \Rightarrow  e^{-x}=-\frac{1}{5}(x-5)
\label{densidadevolumetricadiscreta3}
\end{equation}

A solu\c{c}\~ao da equa\c{c}\~ao da forma $e^{-cx}=a_0(x-r)$ \'e:

\begin{equation}
x=r+\frac{1}{c}W \left( \frac{ce^{-cr}}{a_0} \right)
\label{lambertWfunction}
\end{equation} onde W \'e a fun\c{c}\~ao de Lambert. 
De (\ref{densidadevolumetricadiscreta3}) notamos que $c=1$, $r=5$ e $a_0=-\frac{1}{5}$, utilizando o comando
"lambertw" do software livre "octave" obt\'em-se:

\begin{equation}
x=W(-5e^{-5})+5=4,96511423174428
\label{octave}
\end{equation}

Portanto:

\begin{equation}
\frac{hc}{kT\lambda_{max}}=4,965114232 \Rightarrow \lambda_{max}T=0,201405235\frac{hc}{k}=0,002897768 m.K
\label{lambdamax}
\end{equation}

\begin{figure}[!h]
\begin{center}
 \includegraphics[scale=1.0, bb = 50 50 400 300]{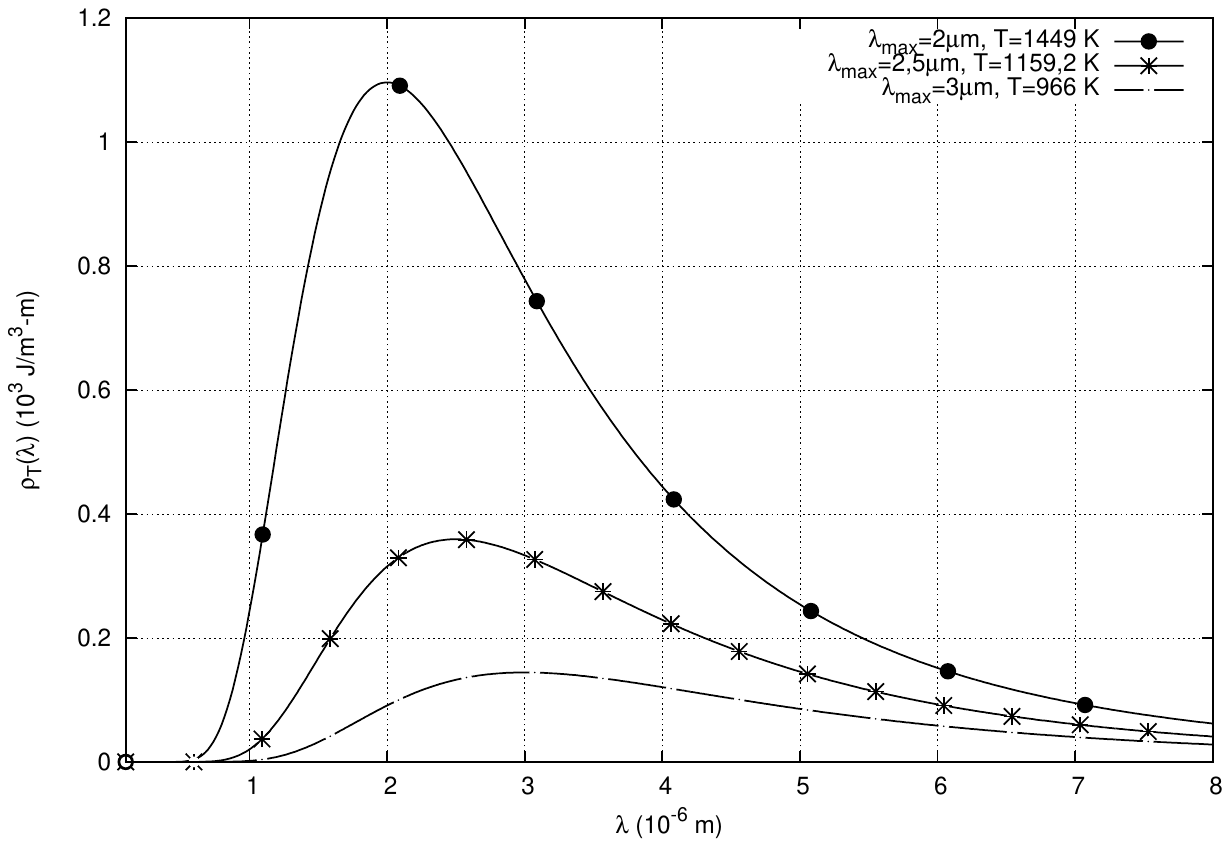}
\end{center}
\caption{Lei de Wien: o valor de $\lambda_{max}$ \'e inversamente proporcional \`a temperatura $T$.}
\label{fig.leidewien}
\end{figure}

\section{Obtendo a lei de Stefan-Boltzmann a partir da f\'ormula da radia\c{c}\~ao de corpo negro de Planck}

Substituindo (\ref{energiatotal4}) em (\ref{RT}):

\begin{equation}
R_T=\frac{c}{4}\int_0^\infty \rho_T(f)df
\label{RT1}
\end{equation}

Substituindo (\ref{densidadevolumetricadiscreta}) em (\ref{RT1}):

\begin{equation}
R_T=\frac{c}{4}\int_0^\infty \frac{8\pi h f^3}{c^3} \frac{1}{exp(\frac{hf}{kT})-1}df
=\frac{c}{4} \frac{8\pi h}{c^3} \int_0^\infty \frac{f^3}{exp(\frac{hf}{kT})-1}df
\label{RT2}
\end{equation}

Fazendo $x=\frac{hf}{kT}$ com $dx=\frac{h}{kT}df$:

\begin{equation}
R_T=\frac{c}{4} \frac{8\pi h}{c^3} \left(\frac{kT}{h}\right )^4 \int_0^\infty \frac{x^3}{e^x-1}dx
\label{RT2}
\end{equation}

Sabendo que $\int_0^\infty \frac{x^3}{e^x-1}dx=\frac{\pi^4}{15}$ (ver se\c{c}\~ao \ref{integralstefanboltzmann}):

\begin{equation}
R_T=\frac{c}{4} \frac{8\pi h}{c^3} \left(\frac{kT}{h}\right )^4 \frac{\pi^4}{15}=\frac{2\pi^5 k^4 h}{15 c^2 h^3} T^4=5,6704 \times 10^{-8} T^4
\label{RT3}
\end{equation}

\newpage

\appendix

\section{Rela\c{c}\~ao entre energia radiada e densidade de energia}
\label{sectionrelacaodensidadeenergia}

Na figura \ref{fig.energiairradiada}, \'e mostrado um elemento de volume $dV$ localizado em $(\theta,\phi,r)$ 
cuja energia atravessa a superf\'icie $A$ centralizada na origem. Vamos considerar um modelo de an\'alise
em que a regi\~ao acima do plano $xy$ \'e o interior da cavidade, a regi\~ao abaixo do plano $xy$ \'e
o exterior da cavidade e a superf\'icie $A$ \'e o orif\'icio de cavidade.

\begin{figure}[!htb]
\begin{center}
 \includegraphics[scale=0.35]{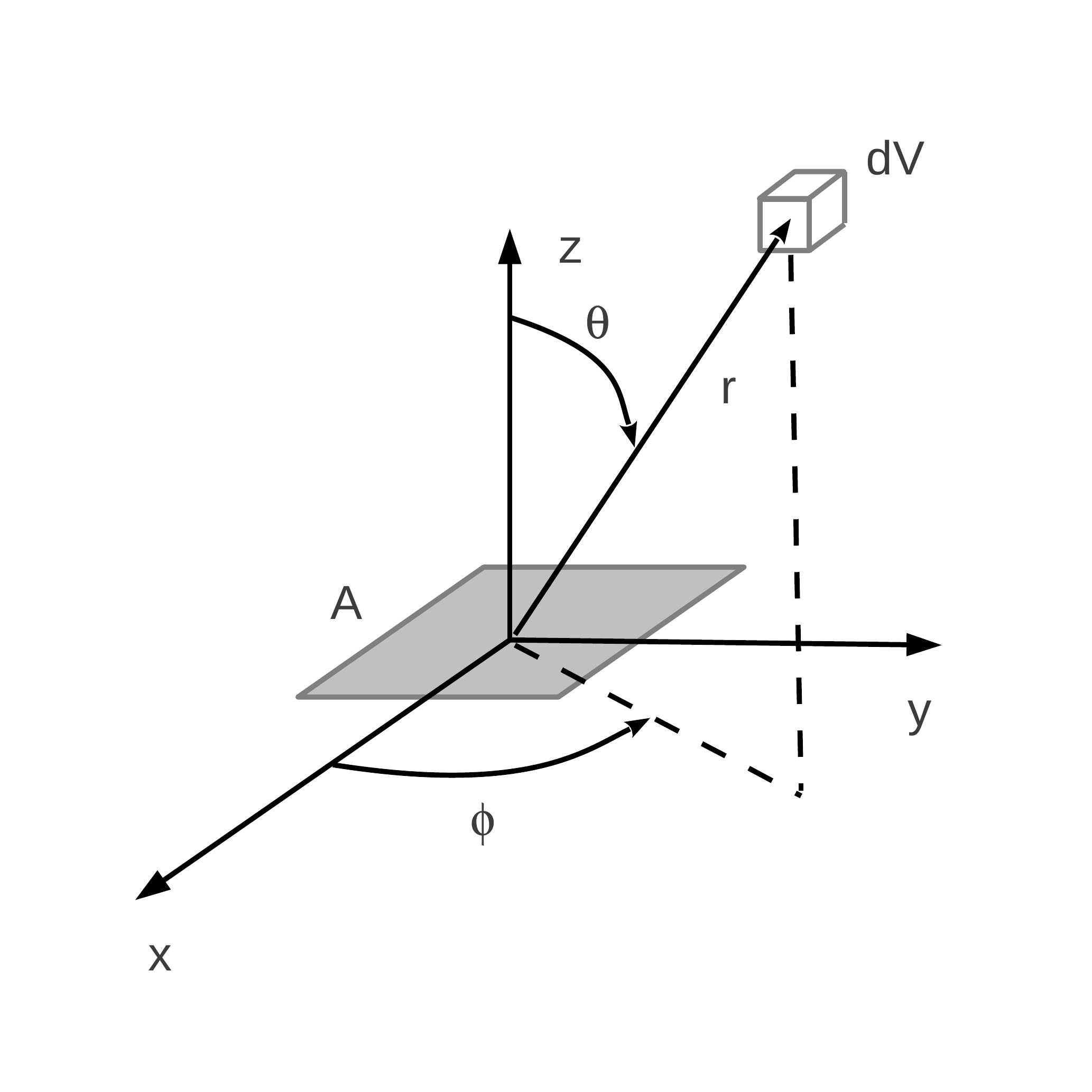}
\end{center}
\caption{Elemento de volume $dV$ localizado em $(\theta,\phi,r)$ irradiando atrav\'es da superf\'icie $A$ centralizada na origem.}
\label{fig.energiairradiada}
\end{figure}

A energia do elemento de volume $dV$ \'e dada pelo produto da densidade volum\'etrica de energia pelo volume $dV$:

\begin{equation}
\rho_T(f)dfdV
\label{contas1}
\end{equation}

A uma dist\^ancia $r$ do elemento de volume $dV$, essa energia est\'a distribu\'ida sobre a superf\'icie de uma esfera, e 
a densidade superficial de energia \'e:

\begin{equation}
\frac{\rho_T(f)dfdV}{4\pi r^2}
\label{contas2}
\end{equation}

A energia $dE_T(f)df$ no elemento de volume $dV$ que passa pela superf\'icie $A$, atravessa uma \'area igual a $A~cos\theta$. 
A energia que passa atrav\'es da superf\'icie $A$ \'e o produto da densidade superficial de energia pela \'area que a radia\c{c}\~ao atravessa:

\begin{equation}
dE_T(f)df=\frac{\rho_T(f)dfdV}{4\pi r^2}A cos\theta
\label{elementoenergia}
\end{equation}

O elemento de volume em coordenadas esf\'ericas \'e:

\begin{equation}
dV=r^2sen\theta d\theta d\phi dr
\label{elementovolume}
\end{equation}

Substituindo (\ref{elementovolume}) em (\ref{elementoenergia}):

\begin{equation}
dE_T(f)df=A \frac{\rho_T(f)df}{4\pi} sen\theta cos\theta d\theta d\phi dr
\label{elementoenergia1}
\end{equation}

Decorrido um tempo $t$, os elementos de volume que contribuem para a energia que passa pela superf\'icie $A$ devem
estar localizados no m\'aximo a uma dist\^ancia $r=ct$ da origem, onde $c=2,99792458\times 10^8 m/s$ \'e 
a velocidade da luz no v\'acuo. Dessa forma, os elementos de volume que dever ser levados em conta est\~ao contidos na
semi-esfera superior de raio $r=ct$, portanto, a energia que atravessa a superf\'icie $A$ em um tempo $t$ \'e:

\begin{equation}
E_T(f)df=A \frac{\rho_T(f)df}{4\pi} \int_{\theta=0}^{\pi/2}\int_{\phi=0}^{2\pi}\int_{r=0}^{ct} sen\theta cos\theta d\theta d\phi dr
\label{energiatotal}
\end{equation}

Integrando em $r$:

\begin{equation}
E_T(f)df=A \frac{\rho_T(f)df}{4\pi} ct \int_{\theta=0}^{\pi/2}\int_{\phi=0}^{2\pi} sen\theta cos\theta d\theta d\phi 
\label{energiatotal1}
\end{equation}

Integrando em $\phi$:

\begin{equation}
E_T(f)df=A \frac{\rho_T(f)df}{4\pi} 2\pi ct \int_{\theta=0}^{\pi/2} sen\theta cos\theta d\theta=
A \frac{\rho_T(f)df}{4\pi} 2\pi ct \left [ \frac{1}{4} \int_{(2\theta)=0}^{\pi} sen(2\theta) d(2\theta) \right ]
\label{energiatotal2}
\end{equation}

Integrando em $\theta$:

\begin{equation}
E_T(f)df=A \frac{\rho_T(f)df}{4\pi} 2\pi ct \frac{1}{2}= A\frac{ct}{4}\rho_T(f)df
\label{energiatotal3}
\end{equation}

Logo:

\begin{equation}
R_T(f)df=\frac{E_T(f)df}{At}= \frac{c}{4}\rho_T(f)df
\label{energiatotal4}
\end{equation}

O que corrobora a hip\'otese levantada em (\ref{RTprop}). Portanto, 
\'e poss\'ivel estudar o problema de irradia\c{c}\~ao por um orif\'icio de cavidade a partir do que ocorre dentro da cavidade.

\section{Densidade de energia}
\label{sectionnumerodemodos}

Uma cavidade ressonante c\'ubica \'e um cubo de paredes condutoras el\'etricamente perfeitas (condutividade el\'etrica 
$\rightarrow \infty$). 
Vamos analisar aqui o caso em que o meio dentro da cavidade \'e o v\'acuo. 
As equa\c{c}\~oes que descrevem os fen\^omenos eletromagn\'eticos s\~ao as equa\c{c}\~oes de Maxwell, as quais na forma fasorial 
e para o v\'acuo, s\~ao:

\begin{eqnarray}
\nabla \times \vec{E}_{s} = - j\omega\mu_0 \vec{H}_s \label{faradayfreq}\\
\nabla \times \vec{H}_{s} = j\omega\varepsilon_0 \vec{E}_s \label{amperefreq}\\
\nabla \cdot \vec{E}_{s} = 0 \label{gausseletrico}\\
\nabla \cdot \vec{H}_{s} = 0 \label{gaussmagnetico}
\end{eqnarray} onde $j=\sqrt{-1}$, $\omega$ \'e a frequ\^encia angular, 
$\varepsilon_0 = 8,85 \times 10^{-12} m^{-3} kg^{-1} s^{4} A^{2}$ \'e a permissividade el\'etrica do v\'acuo, 
$\mu_0 = 1,26 \times 10^{-6} m~kg~s^{-2} A^{-2}$ \'e a permeabilidade magn\'etica do v\'acuo,
e $\vec{E}_s$ e $\vec{H}_s$ s\~ao, respectivamente,
os campos fasoriais el\'etrico e magn\'etico.
 
As equa\c{c}\~oes de onda na forma fasorial s\~ao obtidas atrav\'es da identidade vetorial:

\begin{equation}
\nabla \times (\nabla \times \vec{A}) = \nabla (\nabla \cdot \vec{A})-\nabla^2 \vec{A}
\end{equation}

Para o campo el\'etrico:

\begin{equation}
\nabla \times (\nabla \times \vec{E_s}) = \nabla (\nabla \cdot \vec{E_s})-\nabla^2 \vec{E_s}
\label{identidadevetorialeletrico}
\end{equation}

Substituindo (\ref{faradayfreq}) e (\ref{gausseletrico}) em (\ref{identidadevetorialeletrico}):

\begin{eqnarray}
\nabla \times (-j\omega\mu_0 \vec{H_s}) = -\nabla^2 \vec{E_s} \\
-j\omega\mu_0 \nabla \times \vec{H_s} + \nabla^2 \vec{E_s} = 0 \label{nome1}
\end{eqnarray}

Substituindo (\ref{amperefreq}) em (\ref{nome1}):

\begin{eqnarray}
-j\omega\mu_0 (j\omega\varepsilon_0 \vec{E}_s) + \nabla^2 \vec{E}_s = 0 \\
\omega^2\mu_0\varepsilon_0 \vec{E}_s + \nabla^2 \vec{E}_s = 0 \\
\nabla^2 \vec{E}_s + k^2 \vec{E}_s = 0 \label{ondaeletrica}
\end{eqnarray}

Para o campo magn\'etico:

\begin{equation}
\nabla \times (\nabla \times \vec{H_s}) = \nabla (\nabla \cdot \vec{H_s})-\nabla^2 \vec{H_s} 
\label{identidadevetorialmagnetico}
\end{equation}

Substituindo (\ref{amperefreq}) e (\ref{gaussmagnetico}) em (\ref{identidadevetorialmagnetico}):

\begin{eqnarray}
\nabla \times (j\omega\varepsilon_0 \vec{E_s}) = -\nabla^2 \vec{H_s} \\
j\omega\varepsilon_0 \nabla \times \vec{E_s} + \nabla^2 \vec{H_s} = 0 \label{nome2}
\end{eqnarray}

Substituindo (\ref{faradayfreq}) em (\ref{nome2}):

\begin{eqnarray}
j\omega\varepsilon_0 (-j\omega\mu_0 \vec{H}_s) + \nabla^2 \vec{H}_s = 0 \\
\omega^2\mu_0\varepsilon_0 \vec{H}_s + \nabla^2 \vec{H}_s = 0 \\
\nabla^2 \vec{H}_s + k^2 \vec{H}_s = 0 \label{ondamagnetica}
\end{eqnarray}

Nas equa\c{c}\~oes de onda, (\ref{ondaeletrica}) e (\ref{ondamagnetica}), $k^2=\omega^2 \mu_0\varepsilon_0=\omega^2/c^2$,
onde $c$ \'e a velocidade da luz no v\'acuo. 

Vamos dividir em duas partes o problema de contar os modos, primeiro vamos achar os modos TM em z 
e depois, os modos TE em z. A soma desses resultados d\'a o n\'umero total de modos.

Para obter o n\'umero de modos TM em z devemos fazer $H_{zs}=0$. A partir de (\ref{ondaeletrica}):

\begin{equation}
{{\partial^2 E_{zs}} \over {\partial x^2}} + 
{{\partial^2 E_{zs}} \over {\partial y^2}} + {{\partial^2 E_{zs}} \over {\partial z^2}} + k^2 E_{zs} = 0
\label{ondaEzs}
\end{equation}

Pelo m\'etodo de separa\c{c}\~ao de vari\'aveis:

\begin{equation}
E_{zs}(x,y,z)=X(x)Y(y)Z(z)
\label{EzsXYZ}
\end{equation}

Substituindo (\ref{EzsXYZ}) em (\ref{ondaEzs}):

\begin{equation}
X''(x)Y(y)Z(z)+X(x)Y''(y)Z(z)+X(x)Y(y)Z''(z)+k^2 X(x)Y(y)Z(z)=0
\label{EzsXYZdesenvolvido}
\end{equation}

Dividindo (\ref{EzsXYZdesenvolvido}) por $X(x)Y(y)Z(z)$:

\begin{equation}
{X''(x) \over X(x)} + {Y''(y) \over Y(y)} +{Z''(z) \over Z(z)} + k^2 = 0
\label{EzsXYZdesenvolvido1}
\end{equation}

Fazendo:

\begin{eqnarray}
{X''(x) \over X(x)} =  - k_x^2 
\label{EzsXYZdesenvolvido2a} \\
{Y''(y) \over Y(y)} =  - k_y^2 
\label{EzsXYZdesenvolvido2b} \\
{Z''(z) \over Z(z)} =  - k_z^2 
\label{EzsXYZdesenvolvido2c}
\end{eqnarray}

Resulta:

\begin{equation}
k_x^2+k_y^2+k_z^2=k^2
\label{kxkykz}
\end{equation}

As solu\c{c}\~oes das equa\c{c}\~oes diferenciais ordin\'arias (\ref{EzsXYZdesenvolvido2a}-\ref{EzsXYZdesenvolvido2c}) s\~ao:

\begin{eqnarray}
X(x) = c_1 sen(k_x x) + c_2 cos(k_x x) 
\label{solXYZa} \\
Y(y) = c_3 sen(k_y y) + c_4 cos(k_y y) 
\label{solXYZb} \\
Z(z) = c_5 sen(k_z z) + c_6 cos(k_z z)
\label{solXYZc}
\end{eqnarray}

De (\ref{amperefreq}):

\begin{eqnarray}
{\partial H_{zs} \over \partial y} - {\partial H_{ys} \over \partial z}= j\omega\varepsilon_0 E_{xs} 
\label{ampereA} \\
{\partial H_{xs} \over \partial z} - {\partial H_{zs} \over \partial x}= j\omega\varepsilon_0 E_{ys} 
\label{ampereB} \\
{\partial H_{ys} \over \partial x} - {\partial H_{xs} \over \partial y}= j\omega\varepsilon_0 E_{zs}
\label{ampereC}
\end{eqnarray}

Como $H_{zs}=0$:

\begin{eqnarray}
- {\partial H_{ys} \over \partial z}= j\omega\varepsilon_0 E_{xs} 
\label{ampereAA} \\
{\partial H_{xs} \over \partial z} = j\omega\varepsilon_0 E_{ys}
\label{ampereBB}
\end{eqnarray}

Derivando (\ref{ampereC}) com rela\c{c}\~ao a z:

\begin{equation}
{\partial \over \partial z} ({\partial H_{ys} \over \partial x}) 
- {\partial \over \partial z}({\partial H_{xs} \over \partial y})
= j\omega\varepsilon_0 {\partial E_{zs} \over \partial z}
\label{ampereCC}
\end{equation}

Permutando as derivadas:

\begin{equation}
{\partial \over \partial x} ({\partial H_{ys} \over \partial z}) 
- {\partial \over \partial y}({\partial H_{xs} \over \partial z})
= j\omega\varepsilon_0 {\partial E_{zs} \over \partial z}
\label{ampereCCC}
\end{equation}

Substituindo (\ref{ampereAA}) e (\ref{ampereBB}) em (\ref{ampereCCC}):

\begin{equation}
{\partial \over \partial x} (-j\omega\varepsilon_0 E_{xs}) 
- {\partial \over \partial y}(j\omega\varepsilon_0 E_{ys})
= j\omega\varepsilon_0 {\partial E_{zs} \over \partial z}
\label{ampereCCCC}
\end{equation}

Se $E_{xs}=E_{ys}=0$, ent\~ao:

\begin{equation}
{\partial E_{zs} \over \partial z}=0
\label{newmann}
\end{equation}

Logo, as condi\c{c}\~oes de fronteira em $z=0$ e $z=a$ (onde $a$ \'e a medida da aresta da cavidade c\'ubica) s\~ao:

\begin{equation}
{\partial E_{zs} \over \partial z}=0
\label{newmann}
\end{equation}

E as condi\c{c}\~oes de fronteira em $x=0$, $x=a$, $y=0$ e $y=a$ s\~ao:

\begin{equation}
E_{zs} =0
\label{dirichelet}
\end{equation}

Aplicando a condi\c{c}\~ao de fronteira (\ref{newmann}) em (\ref{EzsXYZ}) para $z=0$ e $z=a$:

\begin{eqnarray}
{\partial E_{zs}(x,y,0) \over \partial z}={\partial (X(x)Y(y)Z(0)) \over \partial z}=
{\partial E_{zs}(x,y,a) \over \partial z}={\partial (X(x)Y(y)Z(a)) \over \partial z}=0 \label{newmann1} \\
{\partial Z(0) \over \partial z}={\partial Z(a) \over \partial z}=0 \label{newmann2} \\
{\partial (c_5 sen(k_z z) + c_6 cos(k_z z)) \over \partial z}=0 |_{z=0} \label{newmann3} \\
c_5=0 \label{newmann4} \\
{\partial (c_6 cos(k_z z)) \over \partial z}=0 |_{z=a} \label{newmann5} \\
k_z=\frac{n\pi}{a},~n=1,2,... \label{newmann6}
\end{eqnarray}

Aplicando a condi\c{c}\~ao de fronteira (\ref{dirichelet}) em (\ref{EzsXYZ}) para $x=0$ e $x=a$:

\begin{eqnarray}
E_{zs}(0,y,z)=X(0)Y(y)Z(z)=
E_{zs}(a,y,z)=X(a)Y(y)Z(z)=0 \label{newmann7} \\
X(0)=X(a)=0 \label{newmann8} \\
c_1 sen(k_z z) + c_2 cos(k_z z)=0 |_{x=0} \label{newmann9} \\
c_2=0 \label{newmann10} \\
c_1 sen(k_z z) =0 |_{x=a} \label{newmann11} \\
k_x=\frac{m\pi}{a},~m=1,2,... \label{newmann12}
\end{eqnarray}

Aplicando a condi\c{c}\~ao de fronteira (\ref{dirichelet}) em (\ref{EzsXYZ}) para $y=0$ e $y=a$:

\begin{eqnarray}
E_{zs}(x,0,z)=X(x)Y(0)Z(z)=
E_{zs}(x,a,z)=X(x)Y(a)Z(z)=0 \label{newmann13} \\
Y(0)=Y(a)=0 \label{newmann14} \\
c_3 sen(k_z z) + c_4 cos(k_z z)=0 |_{y=0} \label{newmann15} \\
c_4=0 \label{newmann16} \\
c_3 sen(k_z z) =0 |_{y=a} \label{newmann17} \\
k_y=\frac{p\pi}{a},~p=1,2,... \label{newmann18}
\end{eqnarray}

De (\ref{kxkykz}), (\ref{newmann6}), (\ref{newmann12}) e  (\ref{newmann18}):

\begin{equation}
k^2=(\frac{m\pi}{a})^2+(\frac{p\pi}{a})^2+(\frac{n\pi}{a})^2
\label{kmpn}
\end{equation}

Como:

\begin{equation}
k^2=\omega^2 \mu_0\varepsilon_0=\omega^2/c^2
\label{komega}
\end{equation}

Ent\~ao:

\begin{equation}
\omega^2=(\frac{c\pi}{a})^2~(m^2+p^2+n^2)
\label{omegampn}
\end{equation}

A equa\c{c}\~ao (\ref{omegampn}) corresponde a uma esfera formada pelos pontos ($m$,$p$,$n$) com raio $r$ 
dado por:

\begin{equation}
r=\frac{\omega a}{c\pi}
\label{raio}
\end{equation}

Como $m\ge 0$, $p\ge 0$ e $n\ge 0$, ent\~ao, todos os modos se encontram no primeiro octante. Para $m<r$, $p<r$ e $n<r$,
o n\'umero de modos \'e $m\times p\times n$ em um volume $m\times p\times n$. 
Portanto, o n\'umero de modos por unidade de volume \'e igual a 1.

Como o volume da esfera no primeiro octante \'e:

\begin{equation}
V(r)=\frac{4\pi r^3/3}{8}=\frac{\pi r^3}{6}
\label{volume1}
\end{equation}

Portanto, o volume compreendido entre $r$ e $r+dr$ \'e: 

\begin{equation}
dV(r)=\frac{dV(r)}{dr}dr=\frac{\pi r^2}{2} dr
\label{volume2}
\end{equation}

Fazendo o produto do n\'umero de modos por unidade de volume pelo volume $dV$ obtemos o
n\'umero de modos TM em z compreendidos entre $\omega$ e $\omega+d\omega$, o qual \'e dado por:

\begin{equation}
1\times dV=\frac{\pi (\frac{\omega a}{c\pi})^2}{2} \frac{d\omega a}{c\pi} = \frac{\pi}{2} (\frac{a}{c\pi})^3\omega^2 d\omega
\label{volume3}
\end{equation}

De forma an\'aloga ao que foi feito nesta se\c{c}\~ao \'e poss\'ivel obter o n\'umero de modos TE em z
(para isso fa\c{c}a $E_{zs}=0$ em vez de $H_{zs}=0$), o qual tamb\'em \'e dado por (\ref{volume3}).

A densidade volum\'etrica de modos dentro da cavidade c\'ubica no intervalo $(\omega,\omega+d\omega)$ \'e:

\begin{equation}
2\times\frac{\frac{\pi}{2} (\frac{a}{c\pi})^3\omega^2 d\omega}{a^3} =  \frac{\omega^2}{c^3\pi^2} d\omega
\label{volume4}
\end{equation}

Escrevendo em termos de $f$ ($\omega=2\pi f$), obt\'em-se
a densidade volum\'etrica de modos dentro de uma cavidade no intervalo de $f$ a $f+df$:

\begin{equation}
\frac{8\pi f^2}{c^3} df
\label{volume}
\end{equation}

Multiplicando a equa\c{c}\~ao anterior pela energia m\'edia $<E>$, 
encontra-se a densidade de energia mostrada em (\ref{densidadedeenergia}). Essa equa\c{c}\~ao tamb\'em pode ser obtida
pela representa\c{c}\~ao da cavidade como um oscilador harm\^onico carregado (ver se\c{c}\~ao 1.1 da p\'agina 3 da refer\^encia
\cite{livro.toledopiza}).

\section{C\'alculo da energia m\'edia}
\label{calculoenergiamedia}

A primeira lei da termodin\^amica para um processo revers\'ivel com varia\c{c}\~ao infitesimal de energia \'e:

\begin{equation}
dU=TdS-PdV
\label{1lei}
\end{equation} onde U \'e a energia, T \'e a temperatura, S \'e a entropia, P \'e a press\~ao e V \'e o volume. Mas:

\begin{equation}
dU=\frac{\partial U}{\partial S}dS+\frac{\partial U}{\partial V}dV
\label{1leiA}
\end{equation}

Portanto:

\begin{equation}
\frac{\partial U}{\partial S}=T
\label{dUdS}
\end{equation}

Ent\~ao:

\begin{equation}
\frac{\partial S}{\partial U}=\frac{1}{T}
\label{dSdU}
\end{equation}

Dados $N$ part\'iculas que podem ser distribu\'idas por $m$ estados distintos, de quantas maneiras diferentes $W$ 
\'e poss\'ivel realizar essa distribui\c{c}\~ao? 

Para isso selecione $N_1$ part\'iculas dentre as $N$ part\'iculas para o estado $1$,
selecione $N_2$ part\'iculas dentre as $N-N_1$ part\'iculas restantes para o estado $2$, 
selecione $N_3$ part\'iculas dentre as $N-N_1-N_2$ part\'iculas restantes para o estado $3$, ... , 
e selecione $N_m$ part\'iculas dentre as $N-N_1-N_2-...-N_{m-1}$ part\'iculas restantes para o estado $m$.  
Portanto:

\begin{equation}
W={N \choose N_1}{N-N_1 \choose N_2}{N-N_1-N_2 \choose N_3}...{N-N_1-N_2-N_3-...-N_{m-1} \choose N_m}
\label{W}
\end{equation}

Desenvolvendo os n\'umeros binomiais:

\begin{eqnarray}
W=\frac{N!}{N_1!(N-N_1)!}\frac{(N-N_1)!}{N_2!(N-N_1-N_2)!}\frac{(N-N_1-N_2)!}{N_3!(N-N_1-N_2-N_3)!} \nonumber \\
\frac{(N-N_1-N_2-N_3-...-N_{m-2})!}{N_{m-1}!(N-N_1-N_2-N_3-...-N_{m-1})!}\frac{(N-N_1-N_2-N_3-...-N_{m-1})!}{N_m!0!}
\label{W1}
\end{eqnarray}

Eliminando os fatores que aparecem tanto no numerador quanto no denominador:

\begin{equation}
W=\frac{N!}{N_1!N_2!N_3!...N_m!}
\label{W2}
\end{equation}

A equa\c{c}\~ao de Boltzmann para a entropia \'e:

\begin{equation}
S=k~ln(W)
\label{S}
\end{equation} onde $k=1,38\times10^{-23} JK^{-1}$ \'e a constante de Boltzmann.

A soma do n\'umero de part\'iculas $N_i$ em cada estado $i$ d\'a o n\'umero total $N$ de part\'iculas:

\begin{equation}
N= \sum_{i=1}^m N_i
\label{N}
\end{equation}

A soma da energia dos part\'iculas $N_iE_i$ em cada estado $i$, de energia $E_i$, d\'a a energia total U:

\begin{equation}
U = \sum_{i=1}^m N_i E_i
\label{U}
\end{equation}

Sunbstituindo (\ref{W1}) em (\ref{S}), a entropia $S$ pode ser escrita em termos das vari\'aveis $N_1,N_2,N_3,...,N_m$:

\begin{equation}
S=k~ln(\frac{N!}{N_1!N_2!N_3!...N_m!})=k [ln(N!)-ln(N_1!)-ln(N_2!)-ln(N_3!)...-ln(N_m!)]
\label{S1}
\end{equation} 

Para valores grandes de $N_i$, vale a aproxima\c{c}\~ao de Stirling:

\begin{equation}
ln(N_i!) = N_i ln(N_i) - N_i
\label{aproxima}
\end{equation} 

Usando essa aproxima\c{c}\~ao podemos reescrever (\ref{S1}):

\begin{equation}
S=k~[ln(N!)-\sum_{i=1}^m(N_i ln(N_i) - N_i)]=k~[ln(N!)+N-\sum_{i=1}^m N_i ln(N_i)]
\label{S2}
\end{equation} 

Vamos maximizar a entropia $S(N_1,N_2,...,N_m)$ mostrada em (\ref{S1}), sujeita \`as restri\c{c}\~oes (\ref{N}) e (\ref{U}),
atrav\'es de multiplicadores de Lagrange:

\begin{equation}
{\partial S \over \partial N_i} + 
\lambda_1 {\partial (N- \sum_{i=1}^m N_i) \over \partial N_i} 
+ \lambda_2 {\partial (U - \sum_{i=1}^m N_i E_i) \over \partial N_i} = 0,~i=1,2,...,m
\label{multiplicadores}
\end{equation} onde $\lambda_1$ e $\lambda_2$ s\~ao os multiplicadores de Lagrange.

Derivando (\ref{S2}) com rela\c{c}\~ao a $N_i$:

\begin{equation}
{\partial S \over \partial N_i} = {\partial (k~[-(N_i ln(N_i) - N_i)]) \over \partial N_i} = -k~[1\times ln(N_i)+N_i\times \frac{1}{N_i}-1]=-k~ln(N_i)
\label{dSdNi}
\end{equation}

Portanto:

\begin{equation}
-k~ln(N_i) + \lambda_1 + \lambda_2 E_i = 0,~i=1,2,...,m
\label{multiplicadores1}
\end{equation}

Multiplicando (\ref{multiplicadores1}) por $N_i$:

\begin{equation}
-k~N_iln(N_i) + \lambda_1 N_i + \lambda_2 N_iE_i = 0,~i=1,2,...,m
\label{multiplicadores2}
\end{equation}

Somando as equa\c{c}\~oes para $i=1,2,...,m$:

\begin{equation}
-k~\sum_{i=1}^m N_iln(N_i) + \lambda_1 \sum_{i=1}^m N_i + \lambda_2 \sum_{i=1}^m N_iE_i=0
\label{multiplicadores3}
\end{equation}

Substituindo (\ref{N}) e (\ref{U}) em (\ref{multiplicadores3}):

\begin{equation}
-k~\sum_{i=1}^m N_iln(N_i) = - \lambda_1 N - \lambda_2 U
\label{multiplicadores4}
\end{equation}

Substituindo (\ref{multiplicadores4}) em (\ref{S2}):

\begin{equation}
S=k~ln(N!)+k~N-k~\sum_{i=1}^m N_i ln(N_i)=k~ln(N!)+k~N- \lambda_1 N - \lambda_2 U
\label{S3}
\end{equation} 

Derivando (\ref{S3}) em rela\c{c}\~ao a $U$:

\begin{equation}
{\partial S \over \partial U}= - \lambda_2
\label{dSdU1}
\end{equation}

Substituindo (\ref{dSdU}) em (\ref{dSdU1}):

\begin{equation}
\lambda_2=-\frac{1}{T}
\label{lambda2}
\end{equation}

Da equa\c{c}\~ao (\ref{multiplicadores1}):

\begin{equation}
N_i = e^{(\frac{\lambda_1}{k} + \frac{\lambda_2 E_i}{k})} = e^{(\frac{\lambda_1}{k})} e^{(\frac{\lambda_2 E_i}{k})},~i=1,2,...,m
\label{N_i}
\end{equation}

Fazendo $A=e^{(\frac{\lambda_1}{k})}$ e substituindo (\ref{lambda2}) em (\ref{N_i}):

\begin{equation}
N_i = A e^{-\frac{E_i}{kT}}
\label{N_i}
\end{equation}

Essa equa\c{c}\~ao significa que o n\'umero de part\'iculas que ocupa um estado cai exponencialmente com a energia desse estado.

Considerando a probabilidade
$P_i$ de encontrar part\'iculas no estado de energia $E_i$ proporcional a $N_i$, temos:

\begin{equation}
P_i = B e^{-\frac{E_i}{kT}}
\label{P_i}
\end{equation}

\section{Limite da s\'erie geom\'etrica}
\label{series}

Dada a s\'erie geom\'etrica:

\begin{equation}
S_n = 1 + q + q^2 + ... + q^n
\label{Sn}
\end{equation}

Multiplicando por $q$:

\begin{equation}
qS_n = q + q^2 + q^3 + ... + q^{n+1}
\label{qSn}
\end{equation}

Subtraindo (\ref{qSn}) de (\ref{Sn}):

\begin{equation}
(q-1)S_n = q^{n+1}-1
\label{qSnmenosSn}
\end{equation}

Portanto:

\begin{equation}
S_n = \frac{q^{n+1}-1}{q-1}
\label{qSnmenosSn}
\end{equation}

Para $0<q<1$, o limite quando $n \to \infty$ \'e:

\begin{equation}
\lim_{n\to\infty} q^{n+1}=0
\label{limninfty}
\end{equation}

Logo:

\begin{equation}
\sum_{m=0}^{\infty} q^m = \frac{1}{1-q}
\label{Sninfinito}
\end{equation}

\section{Limite da s\'erie $\sum_{m=1}^{\infty} mq^m$}
\label{series1}

A reduzida de ordem $n$ da s\'erie $\sum_{m=1}^{\infty}mq^m$ \'e:

\begin{equation}
S_n = q + 2q^2 + ... + nq^n
\label{Sn1}
\end{equation}

Multiplicando por $q$:

\begin{equation}
qS_n = q^2 + 2q^3 + ... + (n-1)q^n +nq^{n+1}
\label{qSn1}
\end{equation}

Subtraindo (\ref{Sn1}) de (\ref{qSn1}):

\begin{eqnarray}
(1-q)S_n = q + (2-1)q^2 + ... +(n-(n-1))q^n - nq^{n+1} \nonumber \\
= q + q^2 + ... + q^n - nq^{n+1}  \nonumber \\
= (1 + q + q^2 + ... + q^n) - (nq^{n+1}+1)  \nonumber \\
= \frac{q^{n+1}-1}{q-1}  - (nq^{n+1}+1) \nonumber \\
= \frac{1-q^{n+1}}{1-q}  - (nq^{n+1}+1) 
\label{qSnmenosSn1}
\end{eqnarray}

Portanto:

\begin{equation}
S_n = \frac{1-q^{n+1}}{(1-q)^2}  - \frac{nq^{n+1}+1}{1-q} 
\label{qSnmenosSn2}
\end{equation}

Para $0<q<1$, o limite quando $n \to \infty$ vale o limite (\ref{limninfty}), mas:

\begin{equation}
\lim_{n\to\infty} nq^{n+1}=\lim_{n\to\infty} n \times \lim_{n\to\infty} q^{n+1} = \infty \times 0 = ?
\label{limninfty1}
\end{equation}

Para solucionar essa indetermina\c{c}\~ao usamos a segunda regra de L'Hospital:

\begin{eqnarray}
\lim_{n\to\infty} nq^{n+1}=\lim_{n\to\infty} \frac{n}{q^{-(n+1)}}= \left[ \frac{\infty}{\infty} \right ] \nonumber \\
= \lim_{n\to\infty} \frac{(n)'}{(q^{-(n+1)})'} = \lim_{n\to\infty} \frac{1}{q^{-(n+1)}ln(q)(-1)}
=\lim_{n\to\infty} -\frac{q^{n+1}}{ln(q)} = 0
\label{limninfty2}
\end{eqnarray}

Substituindo (\ref{limninfty}) e (\ref{limninfty2}) em (\ref{qSnmenosSn2}) quando $n\to\infty$:

\begin{equation}
\sum_{m=1}^{\infty} mq^m = \frac{1}{(1-q)^2}  - \frac{1}{1-q} = \frac{1}{(1-q)^2}  - \frac{1-q}{(1-q)^2}
=\frac{1-(1-q)}{(1-q)^2}=\frac{q}{(1-q)^2}
\label{qSnmenosSn2inf}
\end{equation}

\section{Integral usada na dedu\c{c}\~ao da lei de Stefan-Boltzmann pela f\'ormula de Planck}
\label{integralstefanboltzmann}

Aqui vamos mostrar que para $x>0$ vale a express\~ao:

\begin{equation}
\int_0^{\infty} \frac{x^3}{e^x-1}dx=\frac{\pi^4}{15}.
\label{integralstefanboltzmann1}
\end{equation}


De (\ref{Sninfinito}), quando $0<q<1$, temos:

\begin{equation}
\sum_{m=0}^{\infty} q^m = \frac{1}{1-q}
\label{SninfinitoB}
\end{equation}

Fazendo $q(x)=e^{-x}$, como $q(x)=e^{-x}$ \'e monotonicamente decrescente e $q(0)=1$, ent\~ao, 
$x>0 \Rightarrow q(x)<1$, al\'em disso,
a exponencial $e^{-x}$ s\'o pode assumir valores positivos, assim, $0<q(x)<1$. Dessa forma:

\begin{eqnarray}
\sum_{m=0}^{\infty} (e^{-x})^m = \frac{1}{1-e^{-x}} \nonumber \\
\frac{1}{e^{x}} \times \sum_{m=0}^{\infty} (e^{-x})^m = \frac{1}{e^{x}} \times \frac{1}{1-e^{-x}} \nonumber \\
\sum_{m=0}^{\infty} (e^{-x})^{m+1} = \frac{1}{e^{x}-1} \nonumber \\
\sum_{m=1}^{\infty} (e^{-x})^m=\sum_{m=1}^{\infty} e^{-mx} = \frac{1}{e^{x}-1}
\label{SninfinitoB1}
\end{eqnarray}

Substituindo (\ref{SninfinitoB1}) no primeiro membro de (\ref{integralstefanboltzmann1}):

\begin{equation}
\int_0^{\infty} \frac{x^3}{e^x-1}dx=
\int_0^{\infty} x^3\sum_{m=1}^{\infty} e^{-mx}dx=\int_0^{\infty} \sum_{m=1}^{\infty} x^3e^{-mx}dx
=\sum_{m=1}^{\infty} \int_0^{\infty} x^3e^{-mx}dx
\label{integralstefanboltzmann2}
\end{equation}

Fazendo $u=mx$, ent\~ao $du=mdx$, assim:

\begin{equation}
\int_0^{\infty} \frac{x^3}{e^x-1}dx=\sum_{m=1}^{\infty} \int_0^{\infty} \left(\frac{u}{m}\right)^3e^{-u}\frac{du}{m}
=\sum_{m=1}^{\infty} \frac{1}{m^4} \int_0^{\infty} u^3e^{-u}du
\label{integralstefanboltzmann3}
\end{equation}

Sabendo que $\int_0^{\infty} u^p e^{-u}du=\Gamma(p)=(p-1)!$, onde $p$ \'e um n\'umero inteiro e 
$\Gamma(p)$ \'e a fun\c{c}\~ao gama de $p$, ent\~ao:

\begin{equation}
\int_0^{\infty} \frac{x^3}{e^x-1}dx=\sum_{m=1}^{\infty} \frac{1}{m^4} \Gamma(4)=3!\sum_{m=1}^{\infty} \frac{1}{m^4} 
\label{integralstefanboltzmann4}
\end{equation}

Sabendo que $\sum_{m=1}^{\infty} \frac{1}{m^{2p}}=\zeta(2p)=\frac{(2\pi)^{2p}}{2(2p)!|B_{2p}|}$, 
onde $2p$ \'e um n\'umero inteiro par, $\zeta(2p)$ \'e a fun\c{c}\~ao zeta de Riemann e $B_{2p}$ \'e o n\'umero de Bernoulli ($B_4=-1/30$),
ent\~ao (ver se\c{c}\~ao 5.9 na p\'agina 286 da refer\^encia \cite{livro.arfken}):

\begin{equation}
\int_0^{\infty} \frac{x^3}{e^x-1}dx=3!\zeta(4)=6 \times \frac{\pi^4}{90}=\frac{\pi^4}{15}
\label{integralstefanboltzmann5}
\end{equation}


}
\Teil{Bibliography}{
\addcontentsline{toc}{section}{\refname}
\providecommand{\href}[2]{#2}\begingroup\raggedright\endgroup

\fbox{
\begin{tabular}{l}
  Tiago Carvalho Martins is currently professor at the \textit{Universidade Federal do Sul e Sudeste do Par\'a}, \\
  Marab\'a, Brazil. Dr. Tiago Carvalho Martins is interested in computational physics and optimization\\
  techniques applied to electromagnetic problems. \\
\end{tabular}
}

}
\end{document}